**Learning tapestries: a statistical learning substrate for open chaotic systems measured with error**

**Michael LuValle**

**Department of Statistics, Rutgers University**

**ABSTRACT: The problem of statistical inference for open chaotic systems measured with error is complicated by the interaction of the uncertainty introduced by chaos, and the various sources of random or external variation. Here a method of representing measured data from large open chaotic systems subject to error as collections of threads of plausible pseudo future histories to enable statistical analysis is described. This representation provides asymptotically consistent predictive distributions, for use in developing predictive likelihood methods which:**

1. **provide a framework for variable selection,**
2. **provide a framework for Bayesian updating, so for example 4 season ahead predictions learn naturally as the 3rd season ahead is measured.**
3. **allows examination of conditional scenarios along the future histories for planning purposes.**
4. **allows the ranking of variable, delay combinations with higher signal to noise ratio.**

**The method is tested for learning and variable selection by examining its behavior in predicting 9 years across 4 seasons of climate variables, including local temperature and rainfall measurements at two locations, predicting up to 4 seasons ahead.**

The work of Lalley [1] and Judd[2] pointed out that it is not possible to estimate even the path of a chaotic system if measurement error is unbounded (even gaussian), because for any delay map embedding there will be many paths closer to the observations than the actual path. Thus any predictions are multivalued, converging to a conditional probability distribution as diagrammed in Figure 1, creating a kind of chaotic uncertainty principle for natural chaotic systems. Figure 2 compares estimates of the conditional predictive densities of standardized precipitation for a weather station in Fresno for summer 3,4,5, and 6 seasons ahead for years 1 and 7 of the 9 being predicted, based on climate empirical orthogonal functions (Multivariate Enso Index, MEI[3], Pacific Decadal Oscillation, PDO[4],Arctic Oscilation, AO[5], and Indian Ocean Dipole, IOD[6] ) black and solar cycle data (red) using the methods described here. Each shows this multimodality.

Figure 1

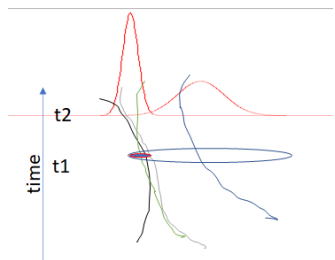

**Legend Figure 1: Here we see four paths of the dynamic system and a measurement is taken at t1. Measurement is at the solid circle. The prediction at t2 is a convolution of the measurement errors and paths possible given the measurement.**

With multimodal densities such as these summaries based on the values of the predictive density at the observed value (predictive marginal likelihoods) are more useful measures of prediction than for example Pearson correlation, because they capture the form of the uncertainty in prediction and not just relationships between means. This is important because much of the current literature uses correlation to choose models for empirical dynamic analysis, which can be misleading about extremes when modes drift apart. Tools for consistent estimation of these predictive marginal likelihoods are readily available and the paragraph below summarizes the argument for their consistency.

**Figure 2**

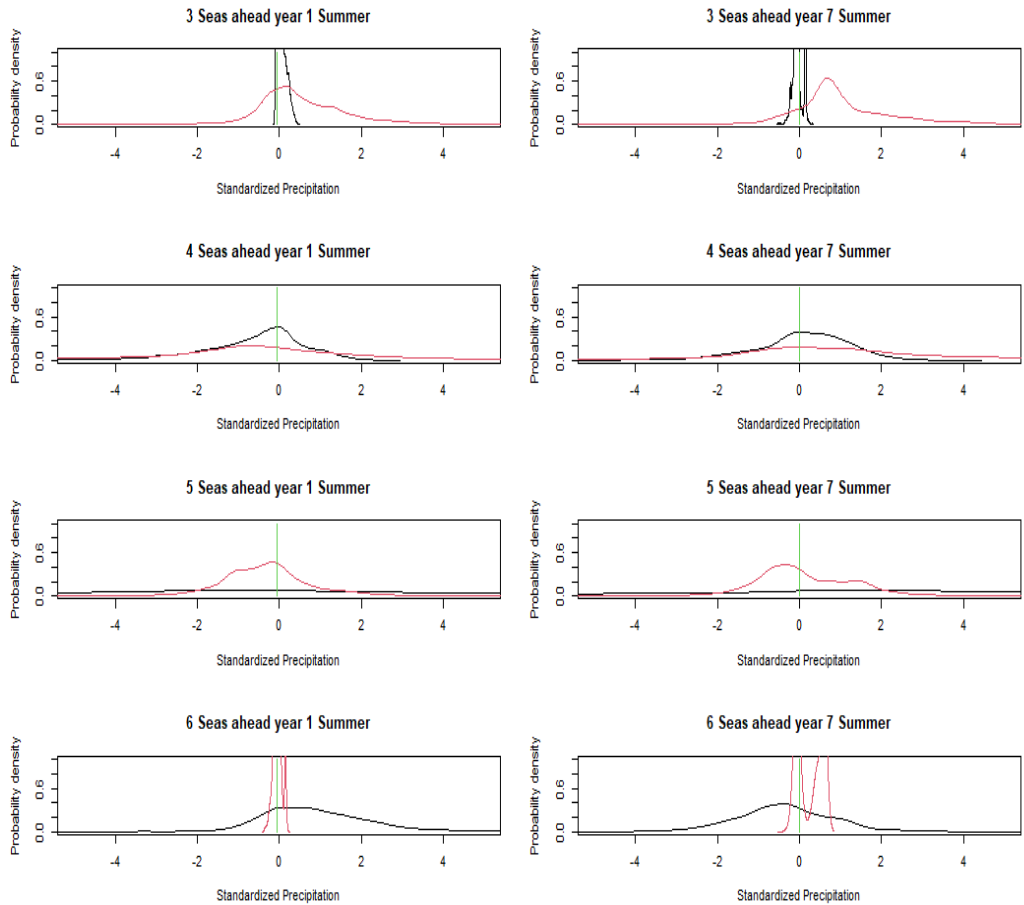

**Legend Figure 2: Here we see 2 predictive densities for summer rainfall in Fresno 3, 4, 5, and 6 seasons ahead for the 1st and 7th year. The observed precipitation is in green, the density in black is constructed from multiview embedding using the four EOF's plus temperature and rainfall, in red from that using the solar cycle data plus temperature and rainfall.**

The framework that can be derived from Lally's work and Judds work is based on an asymptotically consistent estimate of a predictive likelihood for an open chaotic system measured with error using a multiview embedding approach [7,8]. With sufficiently large embedding dimension [9], the set of diffeomorphic embeddings is prevalent [9,10], so local linear regression [11] provides consistent linear prediction for each view within the range of the Lyapunov coefficient for closed chaotic systems with appropriately bounded [1] measurement error. Using Newmann's [12] result combined with Ruelle's [13] on strong mixing of diffeomorphisms of axiom A attractors and an extension of the chaotic hypothesis [14,15], the simple kernal and nearest neighbor density and regression estimates [11] of these predictions based on each view (each embedding is a diffeomorphism), will be consistent. Newmann's result remains valid adding in both measurement error (even gaussian) and small random perturbations in boundary conditions that are jointly strongly mixing with the diffeomorphisms (e.g. independent).

To extend this idea, create a "thread" of future histories to incorporate in a tapestry using a randomly chosen delay map of sufficient estimated size so that if the system was closed it would be a diffeomorphism (for practical purposes, 2.5 times the size estimated by the Levina and Bickel[16] algorithm, as opposed to trying to estimate the box dimension) . This delay map in the data to be predicted defines a point in the training data space, and the j nearest neighbors to that point are found. Then a Y variable k seasons ahead, k-1 seasons ahead, … down to 1 season ahead, is chosen for each trajectory in the nearest neighborhood. For each of the k time gaps to the future a regression is calculated in the training data (the j trajectories in the nearest neighborhood) using a Lars[16] algorithm to select a best regression based on Cp [17]. The same regression model is then applied to the test data, and the predicted Y's have a random residual from the training sample added to it to reflect the contribution of measurement errors.  This creates a series of predictions+errors for 1 up to k seasons ahead. The predictions are all based on the same nearest neighborhood of trajectories collected from a single fixed delay map so internally consistent with one another for a future prediction if the system were measured without error.  This defines a "thread." Thousands of threads are estimated using a random sampling of possible delay maps of the size defined above. The assumption is then that the internal consistency will for some of the threads result in a path consistent with the histories and consistent with the trajectory we are trying to predict. The theoretical conditions under which such threads will exist in sufficient proportion to be useful seem rather tortuous to this author, so this paper examines empirically the hypothesis that  enough of these threads will be consistent with the very low error path so a Bayesian reweighting of the threads as time goes forward by their closeness to observation will result in (stochastic) improvements in predictive likelihood of the resulting reweighted densities of the threads further into the future.

If the hypothesis is true a tapestry of such threads would useful for planning the future by allowing a user (e.g. a water manager controlling a reservoir) to plan how to control their system in a way that takes into account correlations of future extremes like serial drought or flood seasons. So if next season is flooding, should the manager expect flooding, drought, or a normal season the following season given the past. Note that demonstrating that the tapestry can learn in this way implies that it would also be useful in the planning scenario, because learning shows that the associations along the actual path are close to the true interseasonal relationships.

In addition the threads which improve the likelihood by being reweighted can also be identified and examined for variables and lags which are consistent for what is understood about the physics, opening up a new way to probe the physics of systems which are not yet completely understood.

METHOD OVERVIEW: For each predicted season for each variable we can estimate an initial predictive likelihood, by just estimating the density as the derivative of the weighted empirical distribution function in a small neighborhood of the observed value in the test season. As each season appears, we can reweight the threads for following seasons. Here a gaussian function centered at the new observation is used for reweighting the threads as the data comes in. Inference was done using paired t tests on the log predictive likelihoods in sequence through time, using the auto covariance between the log likelihood differences to account for any cross year correlation (likelihoods calculated by predicted season), and multiple tapestry creation to account for the variation between random sampling of delay maps.

      The original data included monthly precipitation and temperature from 2 weather stations (one weather station in Fresno California, the other in volcano national park in Hawaii), the Multivariate Enso Index, the Pacific Decadal Oscillation index, the Arctic Oscillation Index, the Indian Ocean Dipole index, and the number of sunspots. Monthly data was converted to seasonal (averaging temperature and the indeces, and summing precipitation and sunspot number). It was then standardized to create the data sets in R. Within the analysis program the data had their seasonal means subtracted so that the modeling is of the seasonal anomalies.

      The data matrix has 9 columns. For the Fresno matrix, column 1 is Fresno precipitation, 2 is Fresno Temperature, 3 is the MEI, 4 is the PDO, 5 is the AO, 6 is the IOD, 7 is the sunspot cycle, 8 is Hawaii precipitation and 9 is Hawaii Temperature. For the Hawaii matrix, Hawaii precipitation and temperature are 1 and 2, and Fresno 8 and 9 respectively. The run to run variation caused by the random sampling of the delay maps is discussed in the supplementary online information as well as access to the code and data used in this analysis.

      In what follows we will first examine the pointwise difference in likelihood along a tapestry between some different collections of variables that can be used to create predictions. We look both in the natural tapestry format for one tapestry, and, then plots in which each season is pooled together. Finally we look at learning within each tapestry. The primary variable being looked at is rainfall in Hawaii. The data and source software used for this are available on request, so more complete analysis of this data, and any other can be made. False discovery rate analysis is used to emphasize where really significant differences are.

**Empirical Results**

<span style="color:red">Table 1: Spring Tapestry (single replication)</span>

| | Spring | Winter | Fall | Summer |
|---|---|---|---|---|
| 1-9 log likelihood | -12.05437 | -19.37981 | **-6.393650** | -25.02903 |
| 127 log likelihood | **-11.63078** | **-12.54192** | -16.64996 | **-12.64045** |
| P | 0.6804647 | 0.09447211 | 0.000755432 | 3.730899e-07 |
| 1-9 log likelihood | -11.93120 | -19.72020 | **-6.968368** | NA |
| 127 log likelihood | **-11.43007** | **-12.18500** | -16.96903 | NA |
| P | 0.6267011 | 0.10153300 | 0.002122776 | NA |
| 1-9 log likelihood | **-11.49098** | -19.59954 | NA | NA |
| 127 log likelihood | -11.52231 | **-12.29500** | NA | NA |
| P | 0.9710686 | 0.10572820 | NA | NA |
| 1-9 log likelihood | **-11.05210** | NA | NA | NA |
| 127 log likelihood | -11.66909 | NA | NA | NA |
| P | 0.4982969 | NA | NA | NA |

**Legend table 1: Here is the raw log likelihood table generated by 1 tapestry where we are looking at predicting spring 4 seasons ahead. Because the prediction of spring is 4 seasons ahead, we are predicting from a previous spring, so from a software point of view it is efficient to simultaneously predict winter 3 seasons ahead, Fall two seasons ahead and summer 1 season ahead. So the prediction from spring 4 seasons back is represented in the top 3 rows of the table, the first showing**

the multiview predictive log likelihood for the model using all climate variables I have, the second the for the model using only Hawaii precipitation and temperature and the sunspot cycle, and the third, the two sided pvalue for the paired difference t statistic constructed from the two likelihoods measured across 9 years with a covariance calculated using the autocovariance of the 9 pairwise differences. The second 3 rows are the likelihoods calculated with a reweighting of each thread based on how close it's prediction of the measurement 1 season ahead was to the observed measurement, and so on out to the last three rows. We see for example that in this case the spring 1-9 prediction shows improving likelihoods at each step which is at least in the right direction for indicating the Tapestry can learn.

Figure 3 Model comparison across predictions

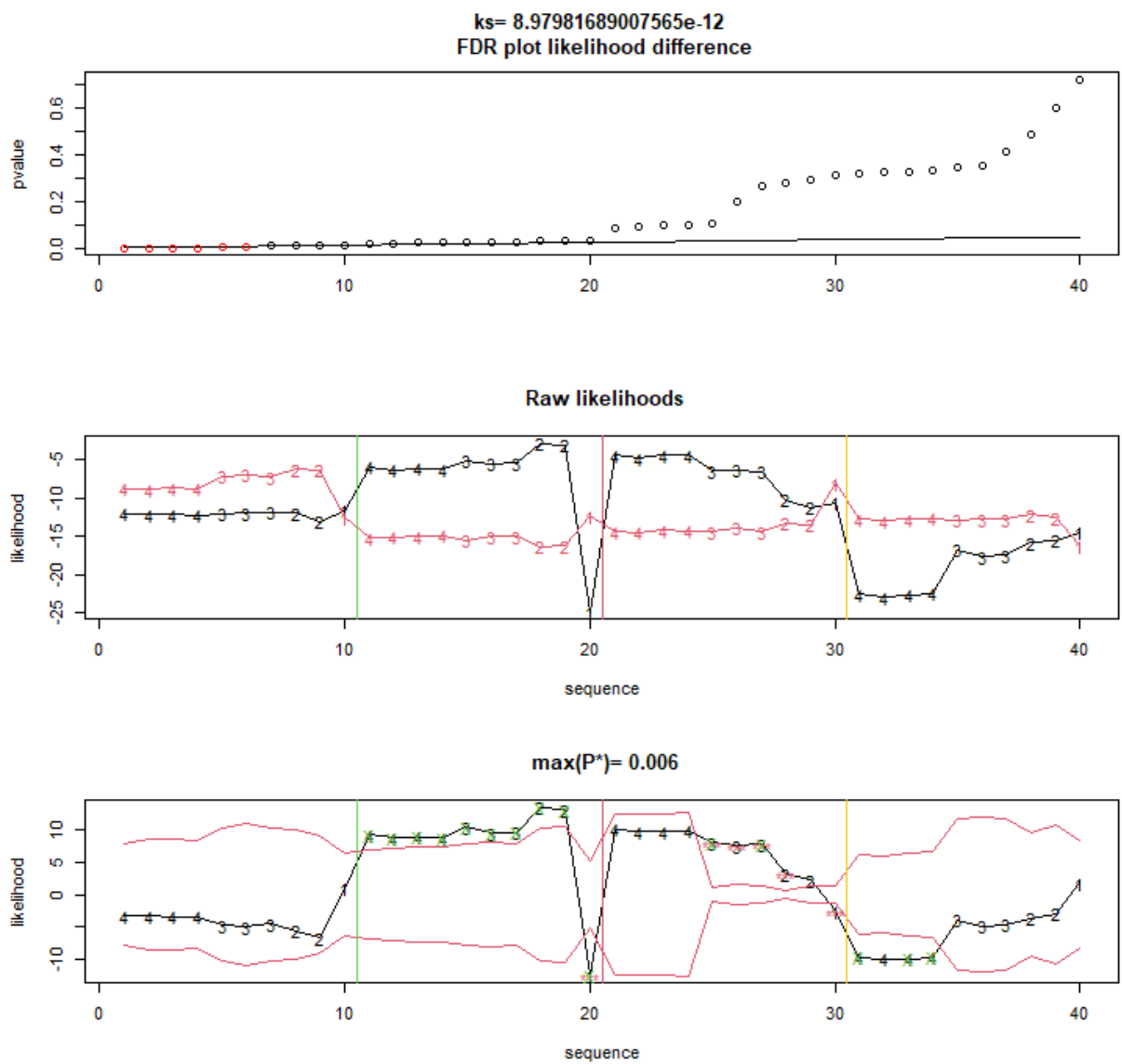

**Legend Figure 3. The plots in figure 3 shows the result across all lags, for both models for Hawaii Precipitation. Here because the results are based on random sampling, 5 runs of each model are used and their mean calculated. The variance used for testing for a difference between the two models is the variance for the mean calculated between runs plus the average variance calculated for the likelihood difference on each run taking into account the autocorrelations between each likelihood difference. The False Discovery Rate plot (at the top of the figure) is for this portmanteau statistical calculation. The second plot shows just the two likelihoods calculated for all the different lags as per table 1. Spring is the first fourth of the table. The four 4 season ahead predictions are plotted first, first that for a full 4 seasons ahead (made right before the beginning of Spring) then the threads are weighted by what happens that summer for the 1st modified 4 season ahead prediction (It is now 3 seasons) then reweighting by fall and finally winter. The three 3 season ahead predictions are next, so from summer to spring, and reweighting again occurs as we get closer, then 2 and finally 1 season ahead all for spring. Then this patter is repeated for summer, fall and winter. The black is the 1-9 (every variable in the delay map) log likelihood, while the red is that using only Hawaiian rainfall, Hawaiian temperature, and the sunspot data. The third plots the difference of the two mean likelihoods around a 95% rejection band, so outside is rejected at a two sided .05 level. Three stars overplot the differences that where found by FDR at a level of .2 (so 20% chance they are false positive) using the adjustment for correlated Pvalues. A green X over plots the data where in all of the individual tapestry calculations the points were identified as positive at an FDR of .2. P\* denotes the largest Pvalue choosen by the FDR evaluation as a possible discovery.**

None of the predictions in spring identify the likelihoods as different, while in summer, most of the full model predictions are identified both by being beyond the 95% test bounds and the within tapestry FDR calculations. However the 1 season ahead prediction is better for the sunspot model by all three criteria. In general we see that particular predictive lags do better for different models which is consistent with what is seen with embedding models in the physics literature (e.g. Bradley and Garland, 2015). Here we verify with extremely noisy data that the differences are statistically significant.

Two data sets are examined for learning, first prediction of Hawaiian precipitation from itself and Fresno temperature and precipitation, shown in figure 4 and prediction of sunspots from the sunspots themselves, and Fresno temperature and precipitation, shown in figure 5. Learning is assessed by an increase in likelihood after reweighting, vs the unweighted likelihood for each lag for prediction.

For leaning in the Hawaiian Precipitation (Figure 4) , there are only 6 points outside the global 95% difference lines in the third plot down. In all four seasons the 1st 3 season ahead prediction shows some learning, In summer both 3 season ahead predictions show a high level of learning, passing both FDR tests. In Fall the 1st 3 season ahead prediction and the 2 season ahead prediction passes the global FDR criterion. In winter both 3 season ahead predictions are beyond the .05 line.

For the sunspot prediction (figure 5), nearly every place learning can occur is being picked as positive by all three criteria, the exception being 3 of the 4 season ahead predictions, although 1 of those is outside the .05 pointwise lines. It is interesting that except for spring, the 1 season ahead prediction based purely on embedding has lower likelihood than any of the other predictions.

**Figure 4: Learning in prediction of Hawaiian Rainfall**

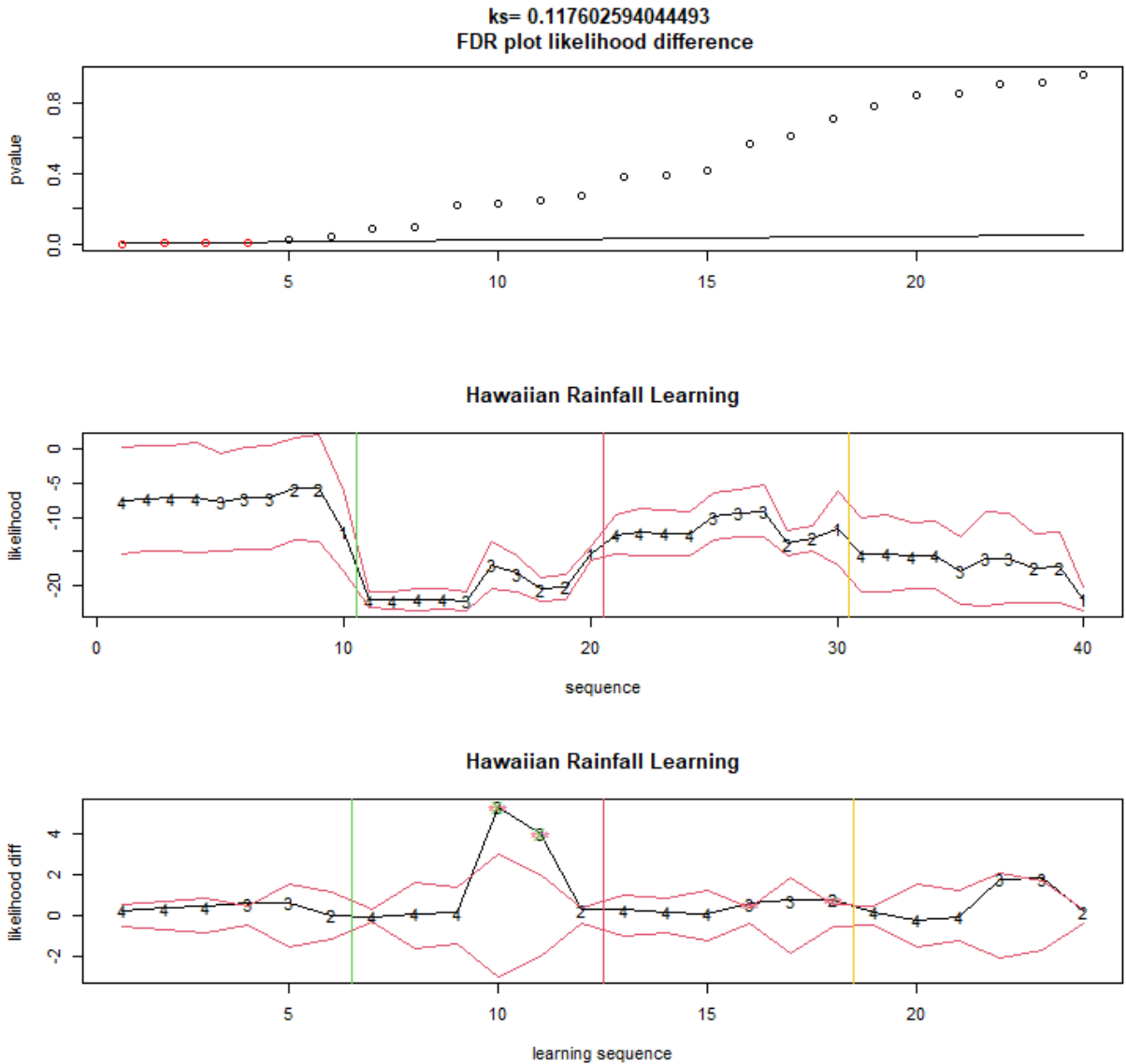

**Legend Figure 4: Here since the likelihood differences are between the first prediction for a given lag and the subsequent predictions for that lag. Otherwise the notation is similar, and the actual likelihood is shown with only 1 line and a simple set of 95% confidence bounds.**

**Figure 5: Learning for sunspot prediction**

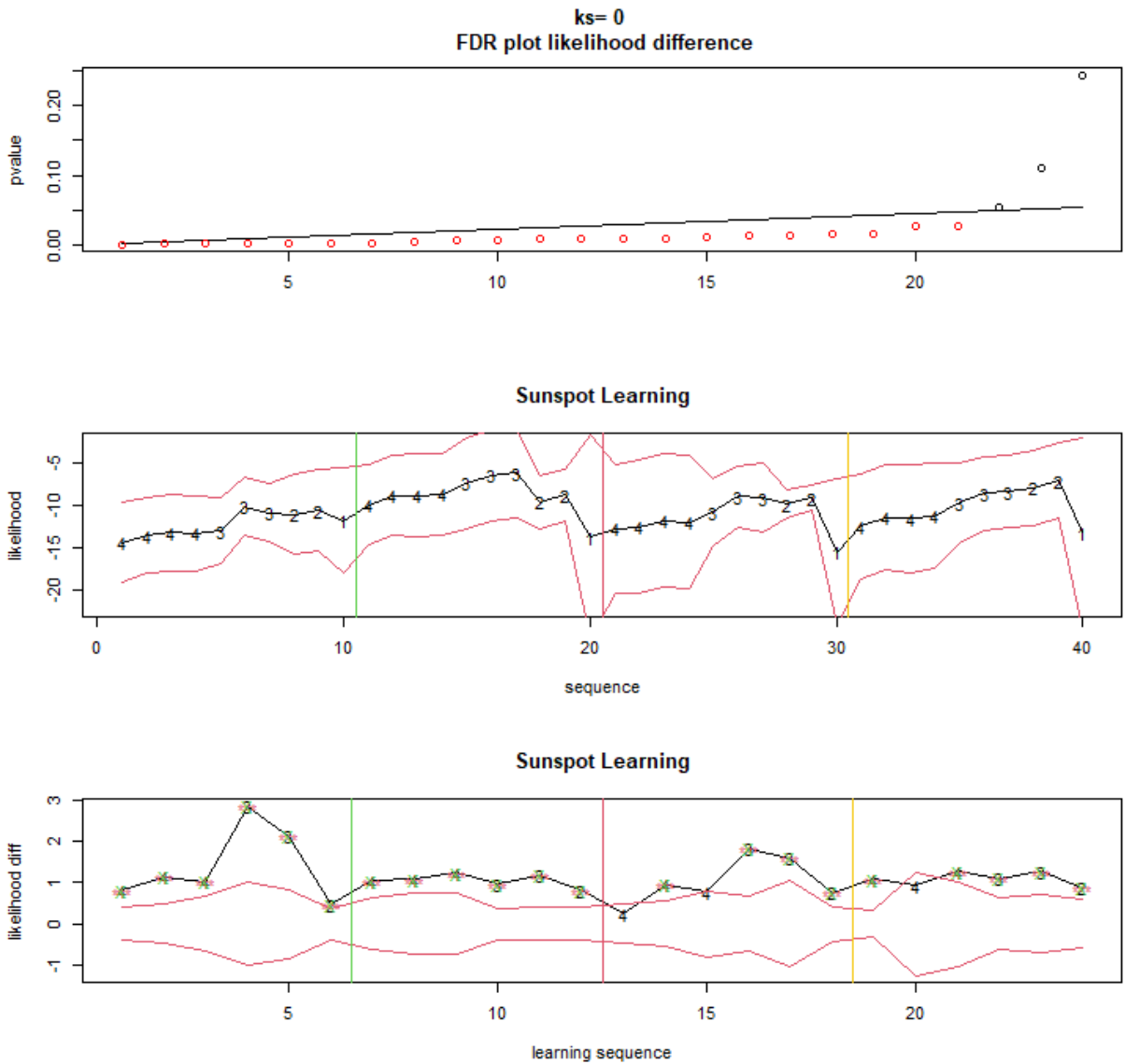

**Legend Figure 5: Here since the likelihood differences are between the first prediction for a given lag and the subsequent predictions for that lag. Otherwise the notation is similar, and the actual likelihood is shown with only 1 line and a simple set of 95% confidence bounds.**

**Discussion**

Both variable selection and learning appear to be working in these data sets. The test data here is only the last 9 years of the data sets, so clearly more validation is necessary. The data and software used to create these results along with a readme describing how to use the programs are available on GitHub, contact the author for access. A large number of further explorations are possible with this data and software alone. For example see if the earth could be used for a probe of the solar cycles see if variables selected from the earth system can result in significantly improve prediction of the sunspots.

More generally this approach could be used to test for any given location whether an empirically based model, a global climate model, or a combination provides better prediction. The method does not depend on the application being climate, any application for which a reasonable generalization of the chaoticity hypothesis holds should be amenable to this approach, for example in confined plasma flow, or more speculatively in flows of social information and activity.

Any commercial application of this approach involving the use of global climate models would have to be done with consultation with my former employer, OFS Laboratories, as they filed and own general patents on my previous work combining climate models with real data for prediction (LuValle, 2016,2019). The author would receive no compensation for such use.

## References and Notes

1) Lalley, 1991, "Beneath the noise, chaos" Annals of Statistics 1999, V27, #2, 461-479
2) Judd, 2007, "Failure of maximum likelihood methods for chaotic dynamic systems", Phys. Rev. E, 75 039210

3) Wolter, K., and M. S. Timlin, 2011: El Niño/Southern Oscillation behaviour since 1871 as diagnosed in an extended multivariate ENSO index (MEI.ext). *Intl. J. Climatology*, **31**, 14pp., 1074-1087. DOI: 10.1002/joc.2336.
4) Mantua, N.J. and Hare, SR (2002), "The Pacific Decadal Oscillation", Journal of Oceanography, Vol. 58, pp. 35 to 44, 2002
5) Higgins, R. W., A. Leetmaa, Y. Xue, and A. Barnston, 2000: Dominant factors influencing the seasonal predictability of U.S. precipitation and surface air temperature. *J. Climate*, **13**, 3994-4017.
6) *Behera, Swadhin K.; Yamagata, Toshio (2003). "Influence of the Indian Ocean Dipole on the Southern Oscillation". Journal of the Meteorological Society of Japan. 81 (1): 169– 177.*
7) Ye, H., and Sugihara G.,(2016), "Information leverage in interconnected ecosystems, overcoming the curse of dimensionality", Science, Vol 353, issue 6502, 922-925
8) Deyle E.R. Sugihara, G (2011), Generalized Theorems for Nonlinear State Space Reconstruction, PLoS ONE 6(3), e18295, doi:10.1371/journal.poine.0018295
9) Sauer, T., Yoreck, J. and Casdagli, M., 1991 "Embedology", Journal of Statistical Physics, 65, 579-616
10) Hunt, Sauer, York, 1992, "Prevalence, a translation invariant "almost every" on infinite dimensional spaces", Bulletin of the AMS, 1992, v27, n2, October 1992, 217-238.
11) Charles J. Stone, *The Annals of Statistics*, Vol. 5, No. 4 (Jul., 1977), pp. 595-620 https://www.jstor.org/stable/2958783
12) Neumann, M.H., (1998), "Strong approximation of density estimators for weakly dependent observation by density estimators from independent observations", Annals of Statistics, V26, no 5, 2014-2048
13) Ruelle, D, (1976), "A measure associated with axiom A attractors", American Journal of Mathematics, V98, no 3, 619-654

14) Bonetto, F., Galavotti, G., Giuliani, A. and Zamponi, F. (2006),"Chaotic Hypothesis, Fluctuation Theorem and Singularities", Journal of Statistical Physics.V123, no 1. DOI: 10.1007/s10955-006-9047-5

15) Galavotti, G. (1996), "Chaotic Hypothesis: Onsager Reciprocity and Fluctuation-Dissipation Theorem", Journal of Statistical Physics, V84, No5/6, 899-925

16) Levina, E. and Bickel, P. J. (2005). "Maximum likelihood estimation of intrinsic dimension". Advances in NIPS 17. MIT Press

17) Trevor Hastie and Brad Efron (2013). Lars: Least Angle Regression, Lasso and Forward Stagewise. R package version 1.2. https://CRAN.R-project.org/package=lars

18) Mallows, C. L. (1973). "Some Comments on $C_P$". *Technometrics*. **15** (4): 661–675. doi:10.2307/1267380. JSTOR 1267380

19) Benjamini and Y. Hochberg, "Controlling the false discovery rate: A practical and powerful approach to multiple testing," J. Roy. Statist. Soc. Ser. B, no. 57, pp. 289–300, 1995.

20) Garland J. and Bradley E., "Prediction in Projection", Chaos **25**, 123108 (2015); doi: 10.1063/1.4936242

21) LuValle, M., 2016, "Predicting climate data using a climate attractors derived from a global climate model" US Patent 9262723, Assignee OFS Fitel

22) LuValle, M., 2019, "Statistical Prediction functions for natural chaotic systems and computer models thereof", Patent 10234595, Assignee OFS Fitel

**Acknowledgments:** I would like to thank Professors Han Xiao, Rong Chen, and John Kolassa for discussion and references to Neumann's article. And I would like to thank Doug Nychka whose comment on an earlier version helped me find a software error that resulted in false regularity in the data. In addition I would like to thank Luke Beebe, Kavi Chakkapi, Yvyn Vyas, Prisha Bhamre, Anastasya Chuchkova and Brian LuValle, whose discussion helped clarify many issues.